# Plasmonic dark field microscopy with a polymer substrate

Yikai Chen, Douguo Zhang[*], Lu Han, Xiangxian Wang, Liangfu Zhu, Pei Wang and Hai Ming

*Institute of Photonics, Department of Optics and Optical Engineering, University of Science and Technology of China, Hefei, Anhui, 230026, China*
*\* Corresponding author: dgzhang@ustc.edu.cn*

In this letter, a plasmonic dark field microscopy taking advantages of the polymer loaded surface plasmon polariton waveguide (PLSPPW) is experimentally demonstrated. The dye molecules (Rhodamine 6G, Rh6G) are doped in the PLSPPW to launch the plasmonic waves or guided waves. Due to the localized property of these waves, the near-field optical energy on the surface of PLSPPW can be scattered into the far-field only in the presence of the objects. The scattering signals then form the dark-field image of the objects. The proposed technique just utilizes a chip-scale integrated plasmonic multilayered structure which is highly compatible with the conventional optical microscopy. The polymer film involved in the PLSPPW also brings about the merits of small roughness, good stability and bio-compatibility. © 2013 Optical Society of America

Dark field microscopy (DFM) is a very simple yet effective technique and well suited for uses involving live and unstained biological samples. Always a beam stop is used in the microscopy to make sure only scattered light by the specimen can reach the detector, which causes the specimen to appear brightly lit against a dark, almost purely black background. The primary advantage to this technique is that it provides high contrast images. With the rapid development of the plasmonic science, DFM demonstrates its powerful function in high contrast optical images and spectra for metallic nanostructures. For examples, high signal to noise ratio scattering spectrum can also be measured with the DFM, which precisely give out the localized surface plasmon resonance (LSPR) wavelength of single gold nanoparticle or its array. On the other hand, the color of metallic structures on the DFM image represents the corresponding LSPR wavelength, so the DFM can also be used to determine the resonant wavelength with high through output[1-3]. However, the high numerical aperture (N.A) condensers used in the conventional DFM are very sensitive to the optical alignment of the set-up, making it difficult to use[4]. Another drawback of the conventional DFM is that it requires high intensity light because of the sharp illumination cone. Thus the conventional DFM is instrumentally complex, costly, and bulky[5]. Recently, a plasmonic DFM utilized a chip-scale multilayer structure is wisely proposed where fluorescence molecules are placed below a thin Ag film to excite the surface plasmon polaritons (SPPs) on the interface between Ag and air[6]. The objects are placed on the Ag film. The dark field images were obtained based on the localized SPPs. Whereas, the exposure of the Ag film to the air made this plasmonic DFM unstable due to the oxidation effect. And also inner properties of the samples may be affected if they are placed on a metallic substrate. To resolve these concerns, in this letter, we propose a new plasmonic DFM taking advantages of recently investigated polymer loaded surface plasmon polariton waveguide(PLSPPW).

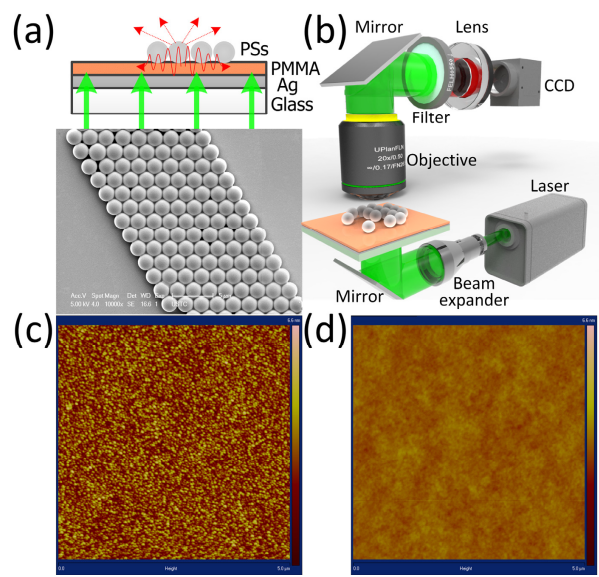

Fig.1. (Color online) (a) the structure diagram of the PLSPPW with PSs on its surface and the SEM image; (b) the schematic diagram of the experimental set-up; (c) and (d), AFM images of the bare Ag film and the PLSPPW.

The structure diagram of the PLSPPW is shown in Figure 1 (a). Dye molecules (Rhodamine 6G, Rh6G) are doped into the PMMA (950K, A2, Micro. Chem) solution which is then spin coated onto a thin Ag film on the glass substrate. The thickness of the PMMA and Ag film are 45nm and 40nm respectively. The multilayer structure is a planar waveguide for the SPPs propagating along the Ag-PMMA-Air interface, which is also called as the PLSPPW[7,8,9]. A two-dimensional hexagonally close packed lattice of the polystyrene spheres (PSs) is formed on top of the PMMA film using a self-assembly method. In the following experiment, the PSs work as the objects to be imaged. The scanning electron microscopy (SEM) image of these PSs on the PLSPPW is shown in Figure 1 (a). The diameter of the PSs is about 2 μm.

Schematic diagram of the experimental setup is shown in Figure 1 (b). A green laser at 532nm wavelength is expanded to illuminate the sample evenly from the glass side. Under illumination, the fluorescence from the Rh6G molecules will launch the SPPs on the Ag-PMMA-Air interface by near field coupling[10, 11]. The SPPs will be scattered by the PSs to the far-field region which are then collected by an objective (20X, N.A 0.6). A long pass filter is put between the objective and a lens to reject the excitation laser beam. Only the fluorescence can be imaged onto the CCD detector. It should be noted that the green laser can be replaced with the light source equipped in the conventional fluorescence microscopy, such as the mercury arc lamp.

Figure 1 (c) presents the atomic force microscopy (AFM, DI Innova, USA ) image of the fresh Ag film on a glass substrate. Figure 1 (d) shows the AFM image of the PLSPPW, which appears softer than Figure 1 (c). The roughness in terms of root mean square of the two images is 0.666nm and 0.160nm respectively. The surface roughness, in addition to the objects, can also scatter the SPPs into free space photons, thus contributing to the background noise. So the better surface flatness of the PLSPPW than that of the bare Ag film is more favorable for the dark-field imaging.

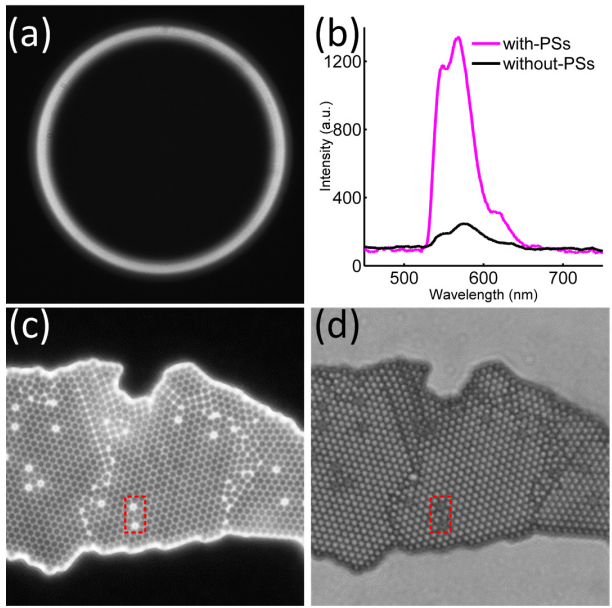

Fig. 2. (Color online) (a) The Fourier plane fluorescence image of the PLSPPW captured by the LRM;. (b) The fluorescence specta from the area with and without the PSs on the PLSPPW; (c ) dark field image of the monolayer PSs; (d) bright field transmission image of the PSs (diamater ~2 μm).

Firstly, the leakage radiation microscopy (LRM) consisted of a high N.A objective (60X, N.A, 1.42) is used to determine the optical modes in the PLSPPW[10, 12]. The Fourier plane image of the LRM is shown in Figure 2 (a), where the PLSPPW is irradiated by an tightly focused green laser beam. The bright ring is the signature of the well known surface-plasmon coupled emission (SPCE), which verifies that SPPs propagating in the PLSPPW are launched by the Rh6G molecules[13,14]. Figure 2 (b) presents the spectra of the fluorescence emitted from the upper side of the PLSPPW captured with the set-up in Figure 1 (b), where the CCD detector is replaced with a spectrometer (Ocean optics, USB 4000). The spectra show that the fluorescence from the area without the PSs is much weaker than that from the area with the PSs. This phenomenon is consistent with the localized property of the SPPs which will be transferred into free space photons only in the presence of scatters. The peak wavelength of the fluorescence locates at 586nm.

The dark field image obtained with the proposed microscopy is presented in Figure 2 (c). The close packed lattice of the PSs displays like the holes array. The other areas without the PSs display black. For each PS, the center part displays black and the edge bright, which is the typical feature of the dark field image. In dark field microscopy, the edges or surfaces of the object scatter light into the far-field which form the DFM image, so the edge looks brighter than the center parts of the object. On the contrary, the center part of the PSs display bright in the bright field transmission image as shown on Figure 2 (d).

We can find some defects in the close packed lattice of the PSs, where there are no PSs, such as the place marked with a red dash-lined box. These defects display black on the bright field image. Whereas, they display much brighter than other PSs on the DFM image. As we know, these defects made more surface of the adjacent PSs exposed to the air, then more light can be scatted into the far-field region. At the center of these defect, there are no object, so there is no light scatted into the far-field. As a result, the black hole with a much bright edge appears on Figure 2 (c). These defects of the lattice are more obviously on the DFM image than on the bright field image.

The comparison of Figure 2 (c) and (d) verifies that dark field image can be realized with the proposed configuration described on Figure 1 (b). The working principle of this DFM can be ascribed to the localization property of the SPPs launched by the excited dye molecules. When the PSs are placed in the near-field of the PMMA film, the SPPs propagating along the Ag-PMMA-Air interface will be scattered into free space photons and then reach the detector, which results in the bright field on the image. On the areas without the PSs, the SPPs cannot radiate into the far-field and result in the black background of the DFM image..

As we know, guided modes can be simultaneously launched in the PLSPPW if the thickness of the PMMA

film is increased, such as the $TM_1$, $TE_1$ modes[15,16]. For these guided modes, the optical field near the PMMA film is evanescent which can be scatted into the far-field by the objects, so it can also be used to realize the dark field imaging. To verify this idea, a PMMA film of 450nm thickness is spin coated on to the Ag film. Figure 3 (a) is the corresponding fluorescence image on the Fourier plane of the LRM. Two bright rings appear which is the evidence of the guided modes launched by the excited dye molecules[15]. Figure 3 (b) is the corresponding DFM image of close packed PSs on the PMMA film. The PS also displays as a hole on the image and the area without the PS display black as the background. So these results confirm the dark field imaging by the guided modes.

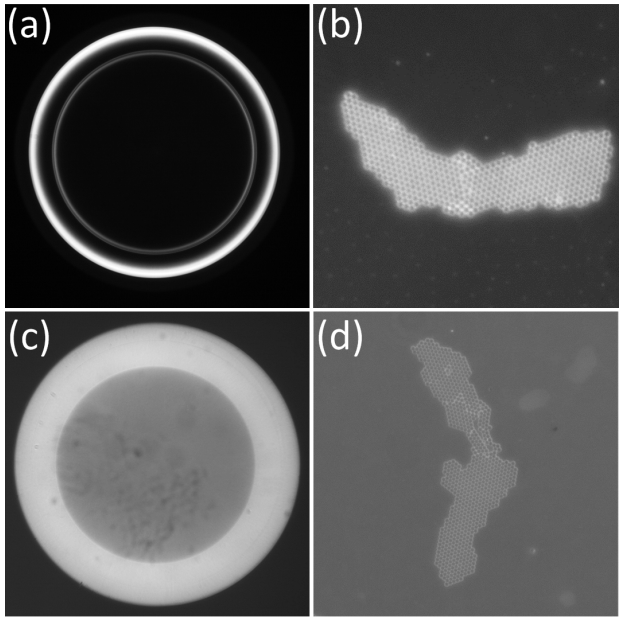

Fig.3. (Color online) (a) The Fourier plane fluorescence image of the PLSPPW with thick PMMA film (450nm) and (b) is the corresponding DFM image of the PSs (diamater ~2 μm); (c) The Fourier plane fluorescence image of the Rh6G doped PMMA film on a glass substrate and (d) is the corresponding DFM image of the PSs on the PMMA film.

Further, evanescent waves can also be generated by the total internal reflection (TIR) at the interface between PMMA and the air. In this case, a Rh6G doped PMMA film is spin coated onto a glass substrate, where there is no Ag film. The excited Rh6G molecules can launch the evanescent wave at the PMMA/Air interface, which is verified by the Fourier plane image of the LRM as shown in Figure 4 (c). There is a wide bright ring appearing on this image, the inner diameter of the ring is corresponding to the critical angle for the TIR, and the outer diameter corresponds to the N.A of the objective (1.42). The fluorescence image of the closed packed PSs on this PMMA film is shown in Figure 3 (d), which also represent the typical feature of the DFM image. So these results verify that the substrate composing of dye molecules doped PMMA film and a glass slip can also realize the dark field imaging.

The difference between the three DFM images is obviously. The ratio of the fluorescence intensity from the point of PS's edge and the point without the PSs (background noise) are 20, 5, and 2.67 for Figure 2 (c), Figure 3 (b) and Figure 3 (d) respectively. The larger the intensity ratio, the better the contrast of the dark field image. In the dark field imaging method, the optical intensity from the area without the objects can be treated as the background noise. In this area, the energy from the excited dye molecules has not coupled into the SPPs, guided modes or the evanescent waves, but radiated into the far field. This difference between the three dark field images demonstrates the following conclusion: for the multilayer film used here, the coupling efficiency from fluorescence to SPPs is higher than that to the guided modes and the TIR.

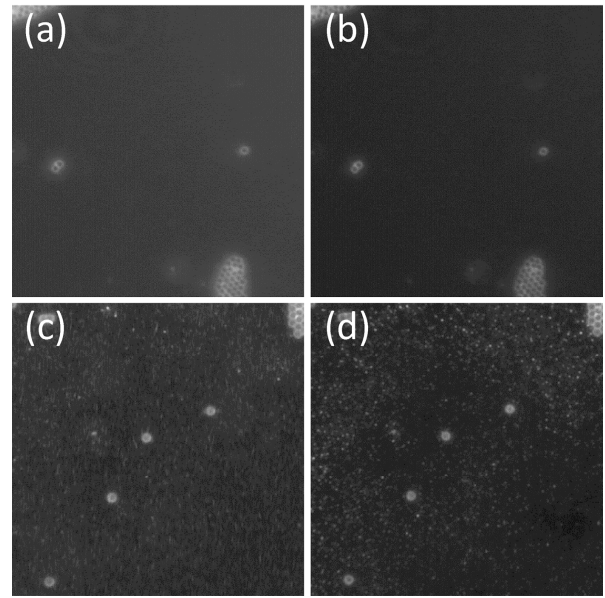

Fig. 4. (Color online) Rh6G doped PMMA film is spin-coated onto an Ag film. (a) (b) are the corresponding dark-field images takedn before and after one week; Ag film is evaporated onto the PMMA film, (c), (d) are the corresponding images taken before and after one week. The diamter of PS is about 2 μm.

At last, the stability of the active PLSPPW enabled DFM is tested. Figure 4 (a) is the dark field image of individual PS on the PLSPPW and (b) is the image of the same area captured after the PLSPPW exposed in the air for one week. For comparison, the multilayer structure composing of the PMMA film below the Ag film is also investigated. Figure 4 (c) is the corresponding dark filed image and (d) is the image taken after one week. In this case, the bare Ag film is exposed in the air. On the four images, the single PS displays a bring ring with a dark center. On Figure 4 (c), there are many bright spots on the area without the PSs which are caused by the roughness of the Ag film and are the background noises of the image. These bright spots become more obviously on Figure 4 (d) due to the oxidation of the Ag which increases the roughness of the Ag film. Whereas, in the case of PMMA film coated on the Ag film, the images look fine where there are no obviously bright spots (a). Figure 4 (b) verifies that this substrate is also stable after one week's exposure in the air. The existence of the PMMA film on the Ag film decreases the surface roughness, which is consistent with the AFM images shown in Figure 1. On the other hand, the PMMA film also protects the Ag film from oxidation and made the dark field imaging technique stable. This is very important in case of long-time observation of the samples ,such as the observation of cell apoptosis.

In summary, a plasmonic DFM with a polymer substrate is experimentally demonstrated here. The dark field imaging can be realized when the objects are placed on a chip-scale multilayer film where SPPs or evanescent waves are launched by the Rh6G molecules doped in the polymer film. The polymer substrate also has the advantages of small roughness, good stability and bio-compatibility. By using the chip-scale multilayer film, a conventional fluorescence microscopy can be easily modified to work as a DFM. The proposed dark field imaging technique has the potential applications in the long time observation of biological samples.

This work is supported by the National Key Basic Research Program of China under grant no. 2013CBA01703, 2012CB921900, 2012CB922003, and the National Natural Science Foundation of China under grant no. 61036005 and 11004182, the Specialized Research Fund for the Doctoral Program of Higher Education (20113402110039), and the Fundamental Research Funds for the Central Universities.

### References

1. W. Andrew Murray and W. L. Barnes, Adv. Mater. 19, 3771–3782 (2007).
2. J. Ye, F.F. Wen,Heidar S, J. Britt Lassiter, P.Van Dorpe, P. Nordlander, and N. J. Halas, Nano Lett. 12, 1660–1667 (2012).
3. J. A. Fan, K. Bao, J. Britt Lassiter, J.M.Bao,N. J. Halas, P. Nordlander,and F. Capasso, ,Nano Lett. 12(6):2817-2821(2012).
4. S. M. Prince and W. G. McGuigan, Proc. SPIE 6676, 66760K (2007).
5. http://en.wikipedia.org/wiki/Dark_field_microscopy
6. Houdong Hu, Changbao Ma, and Zhaowei Liua, Appl.Phys.Lett, 96, 113107(2010).
7. Tobias Holmgaard,and Sergey I. Bozhevolnyi, Phys.Rev.B, 75,245405(2007)
8. Andreas Hohenau, Joachim R. Krenn, Andrey L. Stepanov, Aurelien Drezet, Harald Ditlbacher,Bernhard Steinberger, Alfred Leitner, and Franz R. Aussenegg, Opt.Lett, 30, 893(2005)
9. K. Hassan, A. Bouhelier, T. Bernardin, G. Colas-des-Francs, J.-C. Weeber, and A. Dereux, Phys.Rev.B, 87, 195428 (2013)
10. Douguo Zhang, Xiaocong Yuan, Alexandre Bouhelier. Appl Opt 49(5): 875-879 (2010).
11. Nicolai Hartmann, Giovanni Piredda, Johann Berthelot, Gérard Colas des Francs, Alexandre Bouhelier, Achim Hartschuh. Nano Lett 12(1): 177-181 (2012).
12. A. Drezet,A. Hohenau, D. Koller, A. Stepanov, H. Ditlbacher, B. Steinberger, F.R. Aussenegg, A. Leitner, J.R. Krenn, Materials Science and Engineering: B, 149, 220(2008)
13. J.R.Lakowicz,J.Malicka,I.Gryczynski,Z.Gryczynski,Biochemical and Biophysical Research Communications, 307(3), 435(2003).
14. Douguo Zhang, Qiang Fu, Mingfang Yi, Xiangxian Wang, Yikai Chen, Pei Wang, Yonghua Lu, Peijun Yao, Hai Ming. Plasmonics 7(2): 309-312 (2012).
15. Q. Q. Cheng, T. Li,R. Y. Guo, L. Li, S. M. Wang, and S. N. Zhu, Appl.Phys.Lett, 101,171116(2012)
16. D.G Zhang, X-C Yuan, G. H. Yuan, P. Wang and H. Ming, J. Opt. 12 ,035002,(2010)

# Full citation listings

1. W. Andrew Murray and W. L. Barnes, Plasmonic Materials, Adv. Mater. 19, 3771–3782 (2007).
2. J. Ye, F.F. Wen,Heidar S, J. Britt Lassiter, P.Van Dorpe, P. Nordlander, and N. J. Halas, Plasmonic Nanoclusters: Near Field Properties of the Fano Resonance Interrogated with SERS, Nano Lett. 12, 1660−1667 (2012).
3. J. A. Fan, K. Bao, J. Britt Lassiter, J.M.Bao,N. J. Halas, P. Nordlander,and F. Capasso, Near-Normal Incidence Dark-Field Microscopy: Applications to Nanoplasmonic Spectroscopy,Nano Lett. 12(6):2817-2821(2012).
4. S. M. Prince and W. G. McGuigan, Alignment and Tolerancing of a Cardioid Condenser, Proc. SPIE 6676, 66760K (2007).
5. http://en.wikipedia.org/wiki/Dark_field_microscopy
6. Houdong Hu, Changbao Ma, and Zhaowei Liua, Plasmonic dark field microscopy, Appl.Phys.Lett, 96, 113107(2010).
7. Tobias Holmgaard,and Sergey I. Bozhevolnyi, Theoretical analysis of dielectric-loaded surface plasmon-polariton waveguides, Phys.Rev.B, 75,245405(2007)
8. Andreas Hohenau, Joachim R. Krenn, Andrey L. Stepanov, Aurelien Drezet, Harald Ditlbacher,Bernhard Steinberger, Alfred Leitner, and Franz R. Aussenegg, Dielectric optical elements for surface plasmons, Opt.Lett, 30, 893(2005)
9. K. Hassan, A. Bouhelier, T. Bernardin, G. Colas-des-Francs, J.-C. Weeber, and A. Dereux, Momentum-space spectroscopy for advanced analysis of dielectric-loaded surface plasmon polariton coupled and bent waveguides, Phys.Rev.B, 87, 195428 (2013)
10. Douguo Zhang, Xiaocong Yuan, Alexandre Bouhelier. "Direct image of surface-plasmon-coupled emission by leakage radiation microscopy." Appl Opt 49(5): 875-879 (2010).
11. Nicolai Hartmann, Giovanni Piredda, Johann Berthelot, Gérard Colas des Francs, Alexandre Bouhelier, Achim Hartschuh. "Launching propagating surface plasmon polaritons by a single carbon nanotube dipolar emitter." Nano Lett 12(1): 177-181 (2012).
12. A. Drezet,A. Hohenau, D. Koller, A. Stepanov, H. Ditlbacher, B. Steinberger, F.R. Aussenegg, A. Leitner, J.R. Krenn, Leakage radiation microscopy of surface plasmon polaritons, Materials Science and Engineering: B, 149, 220(2008)
13. J.R.Lakowicz,J.Malicka,I.Gryczynski,Z.Gryczynski, "Directional surface plasmon-coupled emission: a new method for high sensitivity detection", Biochemical and Biophysical Research Communications, 307(3), 435(2003).
14. Douguo Zhang, Qiang Fu, Mingfang Yi, Xiangxian Wang, Yikai Chen, Pei Wang, Yonghua Lu, Peijun Yao, Hai Ming. "Excitation of Broadband Surface Plasmons with Dye Molecules." Plasmonics 7(2): 309-312 (2012).
15. Q. Q. Cheng, T. Li,R. Y. Guo, L. Li, S. M. Wang, and S. N. Zhu, Direct observation of guided-mode interference in polymer-loaded plasmonic waveguide, Appl.Phys.Lett, 101,171116(2012)
16. D.G Zhang, X-C Yuan, G H Yuan, P Wang and H Ming,Directional fluorescence emission characterized with leakage radiation microscopy, J. Opt. 12 ,035002,(2010)